%% file: main.tex
\documentclass[a4paper,11pt]{article}
\usepackage{pos}
\input{package.tex}

\title{Testing (asymptotic) scaling in Yang-Mills theories in the large-$N_c$ limit}
\author*[a,b]{Pietro Butti}
\author[a,c]{Antonio González-Arroyo}

\affiliation[a]{
  Instituto de Física Téorica UAM-CSIC, \\
  Calle Nicolás Cabrera 13-15,
  Universidad Autónoma de Madrid, Cantoblanco, E-28049 Madrid, Spain
}

\affiliation[b]{Departamento de Física Téorica, and
Centro de Astropartículas y Física de Altas Energías (CAPA), \\
Universidad de Zaragoza, Calle Pedro Cerbuna 12, E-50009, Zaragoza, Spain}

\affiliation[c]{
  Departamento de Física Teórica, \\
   Universidad Autónoma de Madrid, Módulo 15, Cantoblanco, E-28049 Madrid, Spain
}

\emailAdd{pbutti@unizar.es}
\emailAdd{antonio.gonzalez-arroyo@uam.es}

\abstract{TEK reduction is a well-established technique that allows single-site simulations of Yang-Mills theory in the large-$N_c$ limit by exploiting volume reduction induced by twisted boundary conditions. We performed simulations for $SU(841)$ for several gauge couplings and applied standard Wilson flow techniques combined with a tree-level improvement methodology to set the lattice scale. The wide range of gauge couplings covered by our simulations allows us to explore a region in the coupling space where our data exhibits asymptotic scaling and perturbation theory could be used to analyze the behaviour of the $\beta$-function. In this talk, I will review the methodology used and go through the main results we obtained, including a determination of the $\Lambda$-parameter of Yang-Mills theory at large-$N_c$ in $\overline{\text{MS}}$-scheme.}

\FullConference{The 40th International Symposium on Lattice Field Theory (Lattice 2023)\\
July 31st - August 4th, 2023\\
Fermi National Accelerator Laboratory\\}

\begin{document}
\begin{flushright}
    \vspace*{-8em}
         IFT-UAM/CSIC-23-58 
    \vspace*{4em}
\end{flushright}
\maketitle

\input{body/0_introduction.tex}

\input{body/1_wilsonflow.tex}
\input{body/2_results.tex}
\input{body/3_conclusions.tex}

\end{document}

%% file: package.tex
\usepackage[T1]{fontenc}    %codifica dei font
\usepackage[utf8]{inputenc} %interpreta i caratteri immessi
\usepackage[english]{babel} %permette di adattare alcuni comandi in italiano
\usepackage{lscape}         % ALLOW PAGE ROTATION

\usepackage{amsmath}        %permette di scrivere formule matemtiche
\usepackage{amsthm}         %permette di inserire teoremi pre-struttrati
\usepackage{amsfonts}		%%fonts for the mathematical rendering of formulas
\usepackage{amscd}          %diagrammi commutativi o di Venn
\usepackage{braket}         %per braket \Braket{"testo del bra"|"testo del ket"}
\usepackage{empheq}         %evidenziare formule
\usepackage{bm}             %lettere greche in grassetto \bm{"lettera"}
\usepackage{physics}
\usepackage{dsfont}         %insiemi numerici fighi \mathbb{robadascrivere}
\usepackage{amsopn}         %dichiarare nuovi operatori
\usepackage{mathrsfs}       %caratteri matematici fighi \mathscr{roba}
\usepackage{xfrac}          %frazioni nel testo \sfrac{num}{den}
\usepackage{bbold}          %numeri tipo la matrice identità \mathbb{1}
\usepackage{slashed}        %notazione di feynmann es: \slashed{\partial}
\usepackage{mathtools}      %per fare le frecce con sopra testo \xrightarrow[testosopra]{testosotto}

\usepackage{booktabs}       %per tabelle
\usepackage{multirow}       %righe multiple in una tabella
\usepackage{multicol}

\usepackage{graphicx}
\usepackage{pgfplots}
    \pgfplotsset{compat=1.13}
\usepackage{tikz}
\usepackage{subfigure}
\usepackage{import}
\usepackage{calc}
\usepackage[font=small,format=hang]{caption} %didascalie

\usepackage[babel]{csquotes}
\usepackage{footmisc}           %per impostare note a pié pagina
\usepackage{cancel}             %per barrare parole e lettere \cancel{paroladabarrare}
\usepackage[nowrite,infront,swapnames]{frontespizio} %per comporre il frontespizio
\usepackage[version=4]{mhchem}  %per formule chimiche, pedici in alto a sinistra
\usepackage{xcolor}
    \definecolor{Goldenrod}{rgb}{0.85, 0.65, 0.13}
    \definecolor{NavyBlue}{rgb}{0.0, 0.0, 0.5}
    \definecolor{BrickRed}{rgb}{0.8, 0.0, 0.0}
    \definecolor{OliveGreen}{rgb}{0.33, 0.42, 0.18}
\usepackage{algorithm}          %Scrivere algoritmi e pseudocodici
\usepackage[all,cmtip]{xy}      %Per fare diagrammini
\usepackage{feynmf}
\usepackage{todonotes}    %per note todo

\usepackage{listings}           %per scrivere codici
    \lstnewenvironment{cppcode}{\lstset{frame=single,language=C++,basicstyle=\small\ttfamily,keywordstyle=\color{blue},commentstyle=\color{darkspringgreen},stringstyle=\color{magenta},backgroundcolor=\color{mygray}}}{}
\usepackage{fancyhdr}
    \fancypagestyle{plain}{
    	\fancyhf{}
    	\rfoot{\thepage}

    }

%% file: body/0_introduction.tex
\section{Introduction}
In a generic bare-coupling definition  scheme labelled by $s$, the Renormalization Group (RG) equation for the bare 't Hooft coupling $\lambda_s=g^2 N_c$ dependence on the cutoff scale $a$ reads
\begin{equation}\label{eqn:RGequation}
    \beta_s(\lambda_s) = - \dv{\lambda_s}{\log a^2}\,.
\end{equation} 
It is well-known that the $\beta$-function has the following perturbative expansion around $\lambda_s\sim 0$
\begin{equation}\label{eqn:betapert}
    \beta(\lambda_s) \sim - b_0 \lambda_s^2 - b_1 \lambda_s^3 - b_2^{(s)} \lambda_s^4 + \order{\lambda_s^5}\,,
\end{equation}
where $b_0$ and $b_1$ are known to be universal (scheme independent) and in a pure Yang-Mills theory they amount to $b_0=\frac{11}{3(4\pi)^2}$ and $b_1 = \frac{34}{3(4\pi)^4}$, while higher-order coefficients are known to be dependent on the scheme chosen. Using a fully perturbative $\beta$-function truncated at $\order{\lambda^3}$, upon integration, Eq.~\eqref{eqn:RGequation} gives
\begin{equation}\label{eqn:aofb}
    -\log{a\Lambda_s} = \frac{1}{2b_0\lambda_s(a)} + \frac{b_1}{2b_0^2}\log(b_0\lambda_s(a))  + \frac{c_1^{(s)}}{2b_0}\lambda_s(a) + \order{\lambda_s^2}\,,
\end{equation}
where the coefficient of the linear term is given by $c_1^{(s)} = \frac{b_2^{(s)}}{b_0} - \frac{b_1^2}{b_0^2}$, and is scheme-dependent, as well as the $\Lambda$-parameter itself. On the lattice, it is natural to use the coupling parameter in the lattice action as a natural coupling scheme. For Wilson action  we label it as $\lambda_w\equiv\sfrac{1}{b}=2 N^2/\beta$. This scheme is known to have large higher-order perturbative corrections, since the ratio between the Wilson and  $\overline{\text{MS}}$ scales is large~\cite{Hasenfratz1980,Gonzalez??} $\frac{\Lambda_{\overline{\text{MS}}}}{\Lambda_w} = 38.853$.
This induces large violations of the scale dependence of the coupling with respect to the truncated perturbative prediction, which we refer to as {\em asymptotic scaling}. This showed up from the earliest studies, and several authors 
proposed adopting a different definition of the coupling which is better behaved. These are called  {\em improved} couplings.  
The goal of this work is to test how well the asymptotic predictions hold for the case  of Yang-Mills theory in the large-$N_c$ limit. Our results are based upon an extensive analysis. Here we present some preliminary results. The full analysis will follow in a future publication~\cite{paper}.

Our strategy is to make use of the volume reduction property at large $N_c$ which allows to obtain results about standard $SU(N_c)$ gauge theory by simulations  on a single site  lattice with twisted boundary conditions~\cite{tek1, tek2}. After a  change of variables, the resulting Wilson action of the reduced model (TEK model) becomes 
\begin{equation}
\label{TEKaction}
    S_\text{TEK} = bN_c\sum_{\mu\neq\nu} \tr\qty[1-z_{\mu \nu} U_\mu U_\nu U_\mu^\dagger U_\nu^\dagger]
\end{equation}
where $U_\mu$ are $SU(N_c)$ matrices, $b=1/\lambda_w$ is the inverse of ’t Hooft coupling $\lambda=g^2N_c$, while
$z_{\mu\nu}$  is a complex phase encoding the twist and given by $e^{i\frac{2\pi k}{N_c}}$ for $\nu>\mu$ ($k$ is an integer  coprime with $\sqrt{N_c}$). This particular choice of the twist factor is called symmetric twist.

The vacuum configurations for the TEK action ~\eqref{TEKaction} are called twist-eaters and they are given by $U_\mu=\Gamma_\mu$ which are  the solutions of the twist equation:
\begin{equation}
    \Gamma_\mu\Gamma_\nu = z_{\nu\mu}\Gamma_\nu\Gamma_\mu
\end{equation}
The equivalence of the TEK model and the infinite volume large $N_c$ gauge theory has been proven both perturbatively and non-perturbatively. Furthermore, the proof shows that finite $N_c$ corrections take the form of finite volume corrections in a lattice whose effective size is $V=(\sqrt{N_c})^4$~\cite{tek1, tek2}.
Thus our choice of $N_c=841$ implies a lattice of size $29^4$ allowing us to study the approach to the continuum limit and matching the expected behaviour of the infinite volume large $N_c$ gauge theory.

%% file: body/1_wilsonflow.tex
\section{The Wilson flow scale(s)}
To set the scale we use an observable which allows a simple determination with great precision. This is the gradient flow scale~\cite{Luscher:2010iy}. This is based upon evolving the gauge field with the equation $\partial_t A_\mu(x,t) = D_\nu G_{\mu\nu}(x,t)$, $t$ being the flow time. As  observable we take  the \text{flowed energy density} $E(t) = \expval{\tr \qty(G_{\mu\nu}(x,t)G_{\mu\nu}(x,t))/2}$. On the single-site twisted lattice the evolution takes place with respect to the dimensionless lattice flow time  $T=\frac{t}{a^2}$, with the corresponding observable being
\begin{equation}\label{eqn:symme}
    E = -\frac{1}{128}\sum_{\mu\neq\nu} \tr[z_{\nu\mu}(
    U_\nu U_\mu U_\nu^\dagger U_\mu^\dagger +
    U_\mu U_\nu^\dagger U_\mu^\dagger U_\nu +
    U_\nu^\dagger U_\mu^\dagger U_\nu U_\mu + 
    U_\mu^\dagger U_\nu U_\mu U_\nu^\dagger)  - \text{h.c.}
    ]^2\,,
\end{equation}
in terms of which we define 
 the dimensionless observable $\Phi(T) = \frac{\expval{T^2E(T)}}{N_c}$
normalized to the number of colours in order to have a finite large $N_c$ limit.
\begin{figure}[t]
    \centering
    % \hspace*{-2cm}
    \includegraphics[scale=0.3]{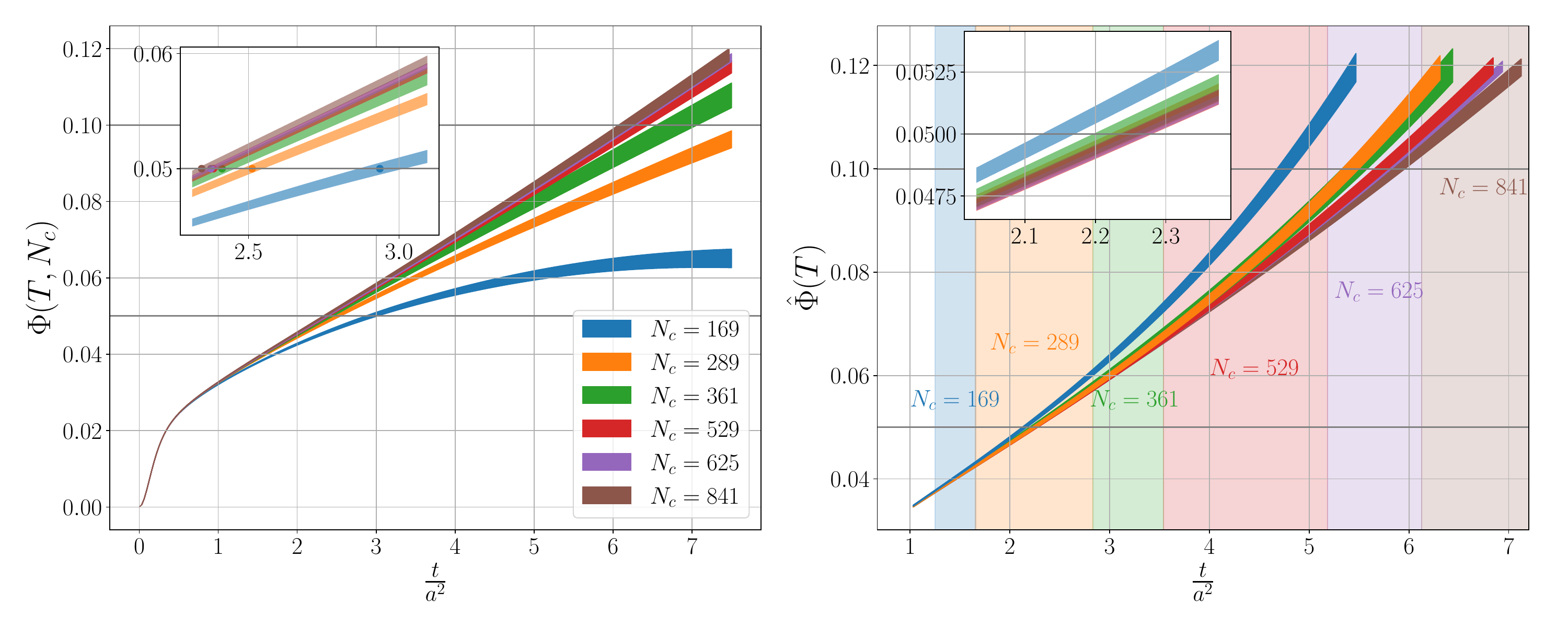}
    \caption{\textit{Left-hand side}: Flowed dimensionless energy density for pure Yang-Mills for different numbers of colours at $b=0.37$. \textit{Right-hand side}: Same flow curves with the norm correction applied. The scaling windows in Eq.~\eqref{eqn:scalingw} are represented with a vertical stripe of the same colour as the flow curve they are associated with. They all start at $T=1.25$ but only the ones with the smallest $N_c$ are depicted so as not to make the colours overlap.}
    \label{fig:flow}
\end{figure}
This observable allows to define a scale as follows
\begin{equation}\label{eqn:scaleeq}
    \eval{\Phi(T)}_{T=T_1} \equiv \frac{1}{N_c}\eval{ \expval{T^2 E(T)}}_{T=T_1} = 0.05.
\end{equation}
The choice of $0.05$ is taken as a compromise that minimizes finite volume effects (finite $N_c$) while still being very slightly affected by lattice artefacts. To exemplify this, we display in the left side of Fig.~\ref{fig:flow} the flow curves for several values of $N_c$ for pure Yang-Mills reduced model ($N_f=0$) at $b=0.37$, where the two reference scales $0.1$ and $0.05$ are depicted with corresponding horizontal grey lines. As it is visible from the plot, although the scale $T_1$ is less affected by the $N_c$-dependence of the flow, the systematic error induced is still sizeable for the smaller values of $N_c$. The situation can be strongly ameliorated by the method employed in Ref.~\cite{Butti:2022sgy} to be briefly explained below. 

Our approach is to consider a new lattice observable to replace $\Phi(T)$, which nonetheless coincides with its infinite $N_c$ and vanishing lattice spacing limit. It is clear that this quantity is as good an observable as the previous one, but is expected to have smaller finite $N_c$ and lattice corrections.  This is given by 
\begin{equation}
    \hat\Phi(T) = \frac{3}{128\pi^2\hat{\mathcal{N}}_N(T)}\Phi(T,N)
\end{equation}
where the prefactor $\hat{\mathcal{N}}_N(T)$ is a function 
of $\sqrt{\sfrac{8T}{N_c}})$ such that 
 $\hat{\mathcal{N}}_\infty(0)=\frac{3}{128\pi^2}$. It is determined in such a way as to eliminate the $N_c$ dependence at the lowest order of perturbation theory. This was analytically calculated in Ref.~\cite{flow}.  We refer to our modified method as \textit{norm correction} and we expect it to simultaneously reduce finite-$N_c$ and lattice artefacts in an effective window in $T$. The lower limit is set to eliminate any interference from lattice artefacts, while the upper limit is determined by ensuring that the smearing radius remains reasonably smaller than a fraction of the overall size of the effective lattice, where residual finite-volume effects are negligible. We typically refer to this range in $T$ as the \textit{scaling window}, which can be expressed as
\begin{equation}\label{eqn:scalingw}
    \qty[1.25,\gamma^2 \frac{N_c}{8}] \text{ with }\gamma=0.22
\end{equation}
After the application of the norm correction, we define $T_1$ in the same way but in terms of $\hat\Phi(T)$, and we extract the scale $T_1=t_1/a^2$ by interpolation. The right-hand side of Fig.~\ref{fig:flow} shows the effect.
We observe that, apart from the case of $N_c=169$, all the scales $T_1$, interpolated from $\hat\Phi$, coincide within errors, signalling that the norm-correction can effectively correct the finite-$N_c$/finite-size effects. The norm-correction method was used in Ref.~\cite{Butti:2022sgy} to determine the scale from data of $N_c=289$ and $361$. In our case, since $N_c=841$ is much larger we can safely claim that the corresponding scales amount to the corresponding one at infinite $N_c$. A detailed discussion about the norm correction and its applications to the case of this and other theories can be found also in\cite{pietrothesis}.

We have measured the scale $T_1$ for a large number of values of $b$ for the TEK model.  The obtained values of $1/\sqrt{8 T_1}$  are listed in the second column of  Tab.~\ref{tab:ratio}.
\begin{table}
    \centering
    \begin{tabular}{lll}
        \toprule
            $b$ & $\frac{a}{\sqrt{8t_1}}$ & $a\sqrt{\sigma}$ \\
        \midrule
            $0.355$  &  $0.37689(62)$ &  $0.2410(30)$ \\
            $0.360$  &  $0.31754(47)$ &  $0.2058(25)$ \\
            $0.365$  &  $0.2716(10) $ &  $0.1784(17)$ \\
            $0.370$  &  $0.23582(69)$ &  $0.1573(19)$ \\
            $0.375$  &  $0.20605(69)$ &  $0.1361(17)$ \\
            $0.380$  &  $0.17734(86)$ &  $0.1191(17)$ \\
            $0.385$  &  $0.1584(15) $ &  $0.1049(11)$ \\
        \bottomrule
    \end{tabular}
    \caption{Lattice scale in units of the Wilson flow improved scale and of the string tension (taken from~\cite{Gonzalez-Arroyo:2012euf}).}
    \label{tab:ratio}
\end{table}

%% file: body/2_results.tex
\begin{figure}
    \centering
    \includegraphics[scale=0.45]{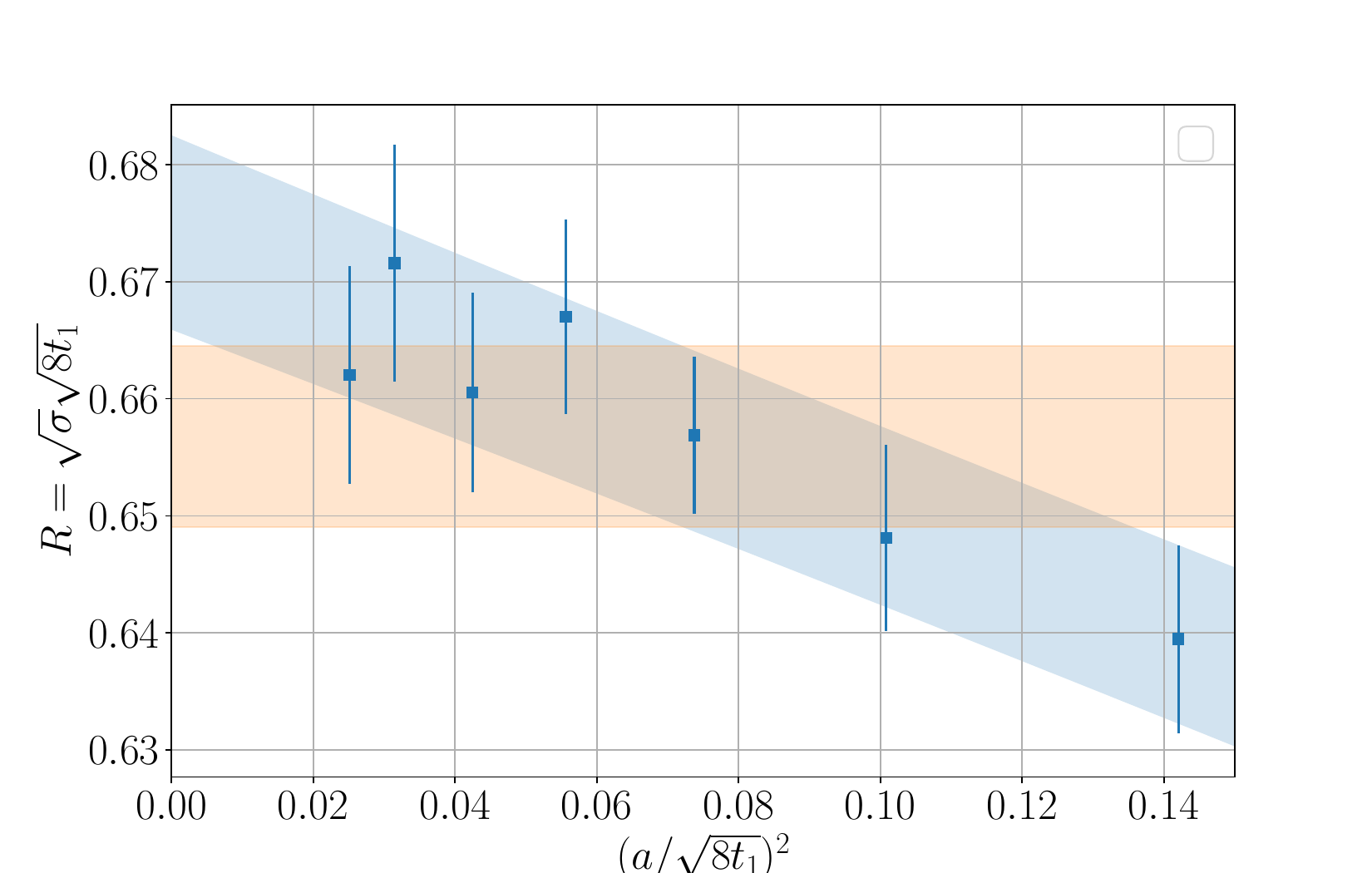}
    \caption{Ratio of the string tension on the lattice and the Wilson flow scale for each value of the gauge coupling $b$. The horizontal band corresponds to the fit to a constant, while the linear fit gives a value of $0.6742(83)$ for the extrapolated value.}
    \label{fig:ratio_sigma}
\end{figure}

We can confront the extracted scales with a previous, less precise,  determination in units of the string tension $\sigma$ given in  Ref.~\cite{Gonzalez-Arroyo:2012euf}.  The numbers are reported in the third column of Tab.~\ref{tab:ratio}. In Fig.~\ref{fig:ratio_sigma}, we depict each value of the dimensionless ratio $R=\sqrt{\sigma}{\sqrt{8t_1}}$ as a function of $(a/\sqrt{8t_1})^2$. 
Exact scaling would imply the ratio to be constant. Such a fit has a  $\chi^2/\#   \mathrm{dof}=1.6$, mainly spoiled by the value at the 2 coarsest point. Nevertheless, we notice that a linear fit in $(a/\sqrt{8t_1})^2$ gives an $\chi^2/\#\text{dof}=0.24$ and might suggest slight  scaling violations. As a final value, we will consider $0.674(8)(18)$ where the systematic error is the dispersion between the two previously obtained values. This comparison with the string tension is a remarkable confirmation of the validity of scaling for a wide range of ’t Hooft couplings with only small violations in the case of the coarsest ensembles.
% In the next subsection, we will exploit this feature to extract the $\Lambda$-parameter of the Yang-Mills theory in the large-$N_c$ limit. We recall that what is going to be presented in the next sections is the result of a preliminary analysis, and there is ongoing work that is not included. The final results will be presented in a future publication\textcolor{red}{add ref}.

\section{Asymptotic scaling and $\Lambda_{\overline{\text{MS}}}$}
\begin{figure}
    \centering
    \includegraphics[scale=0.45]{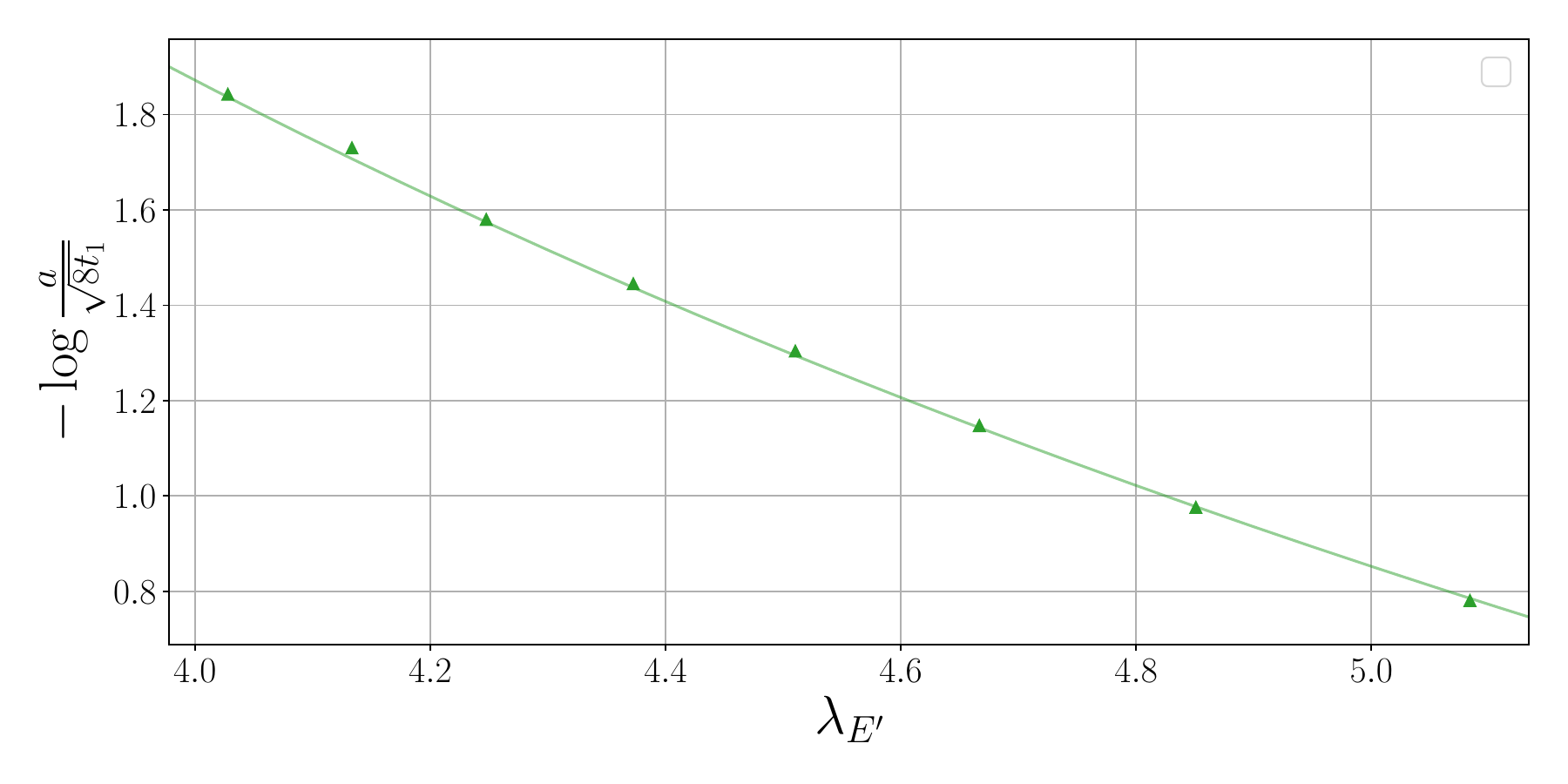}
    \caption{Logartihm of the lattice spacing in units of the Wilson flow scale $\sqrt{8t_1}$ listed in~\ref{tab:ratio} as a function of the improved coupling $\lambda_{E'}$ defined in Tab.~\ref{tab:improvedcouplings}. The green line is the result of a fit to Eq.~\eqref{eqn:aofb}.}
    \label{fig:a_vs_b}
\end{figure}

In this section, we will analyze if the scales we determined follow the asymptotic predictions reviewed in the introduction. For the large $N_c$ theory a similar analysis was performed earlier ~\cite{Allton:2008ty, Gonzalez-Arroyo:2012euf}. We employ three different improved couplings: $\lambda_I$ (from~\cite{Allton:2008ty}), $\lambda_E$ (from~\cite{Martinelli:1980tb,Edwards:1997xf}) and $\lambda_{E'}$, defined in Tab.~\ref{tab:improvedcouplings}.
\begin{table}[t]
    \centering
    \begin{tabular}{cccc}
        \toprule
            symbol         & definition              & $\frac{\Lambda_w}{\Lambda_s}$ & $c_1$ \\
        \midrule
            $\lambda_w$    &   $\frac{1}{b}$                    &  $1$             &  $-0.00438798235$  \\
        \midrule
            $\lambda_I$    &   $\frac{\lambda_w}{P(\lambda_w)}$ &  $0.0677656398$  &  $-0.00172791$  \\
            $\lambda_E$    &   $8(1-P(\lambda_w))$              &  $0.4148791463$  &  $-0.0005030$   \\
            $\lambda_{E'}$ &   $-8\log P(\lambda_w)$            &  $0.1080025976$  &  $-0.00042426$  \\
        \bottomrule
    \end{tabular}
    \caption{Possible definitions for the gauge coupling on the lattice.}
    \label{tab:improvedcouplings}
\end{table}   
% which we are going to briefly review here. Given a change in the coupling
% \begin{equation}\label{eqn:implambda}
%     \lambda' = \lambda_s(1+\gamma_1\lambda_s + \gamma_2\lambda_s^2) + \order{\lambda_S^4}\,,
% \end{equation}
% , we can easily relate the definition of the scheme dependent quantities (the $\Lambda$-parameter and $c_1^{(s)})$ with a one-loop calculation. In the following, 
% 
% By employing these three different improved couplings, in the following, we proceed with the extraction of the corresponding $\Lambda$-parameter.
% 
Having at our disposal such a large range of values for the 't Hooft couplings and given the small errors of our scale determination we are certainly making stringent tests to our data. A global analysis can be done by fitting Eq.~\eqref{eqn:aofb} to our data for the case of the $E'$ scheme only leaving free the value of $\Lambda_{E'}$. The result is displayed in Fig.~\ref{fig:a_vs_b} where the data points and the continuous best-fit curve are shown. The overall impression is very good since the curve follows quite nicely the evolution of the data. However, given the small errors, the $\chi^2$ of the fit is not good. Indeed, for the range of scales covered, certain non-perturbative scaling violations are to be expected. 

\begin{figure}
    \centering
    \includegraphics[scale=0.4]{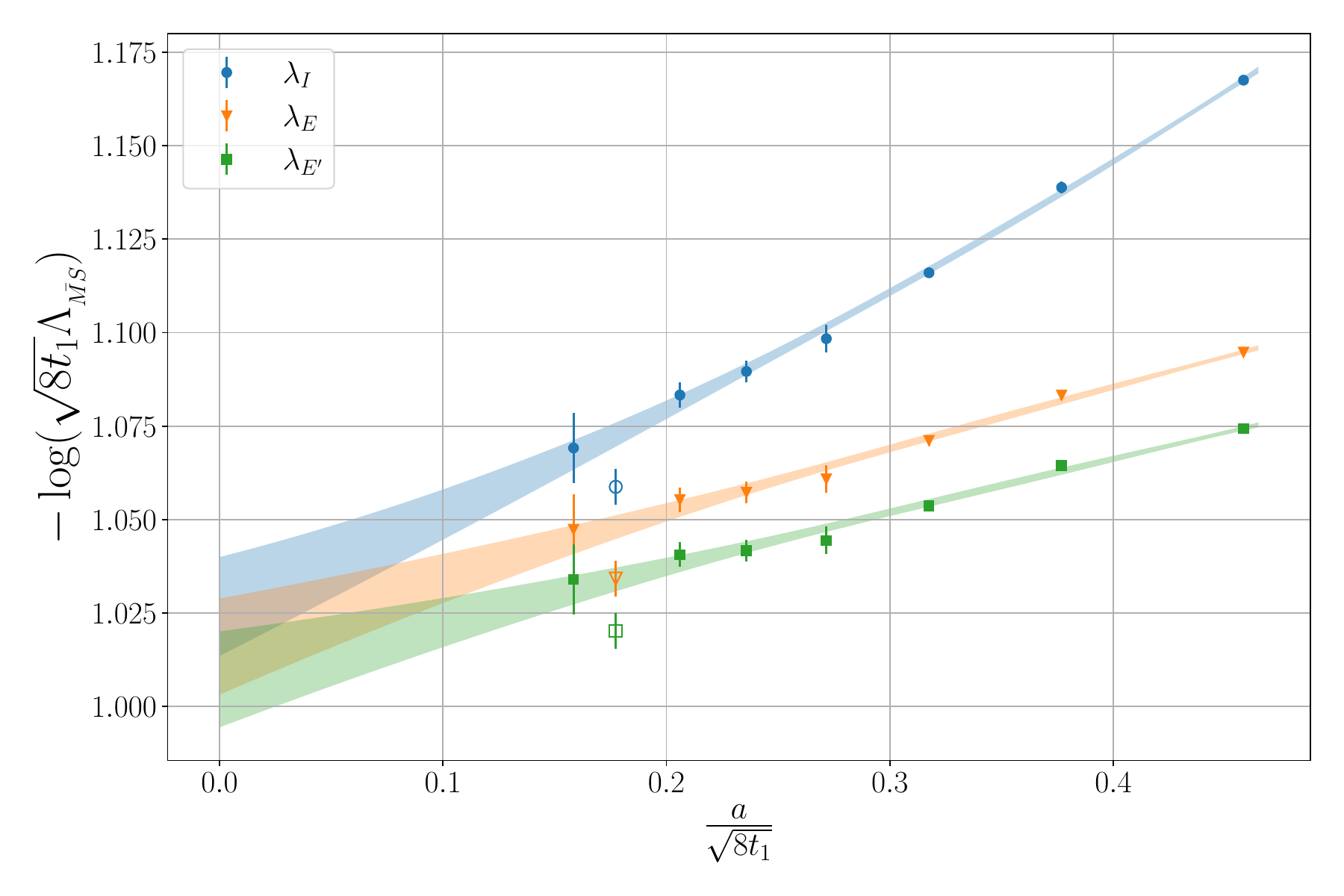}
    \caption{(Logarithm of the) $\Lambda$-parameter in the $\overline{\text{MS}}$ scheme in units of $\sqrt{8t_1}$ versus the lattice spacing in the same units using three different improved couplings in Tab.~\ref{tab:improvedcouplings}. We also depict the corresponding quadratic continuum extrapolations. In the fit, we excluded the point corresponding to $b=0.38$ (whose points are depicted with an empty marker), obtaining $\chi^2/\#\text{dof}=0.49,0.61,0.62$ for $\lambda_I$, $\lambda_E$ and $\lambda_{E'}$, respectively.}
    \label{fig:ascaling}
\end{figure}

We aim at a determination of the $\Lambda$ parameters which are consistent with the predictions of perturbation theory. It is customary to choose as preferred scale the one coming from  the $\overline{\text{MS}}$ scheme in the continuum: $\Lambda_{\overline{\text{MS}}}$.
% 
% where we defined
% \begin{equation}
%     f_s(\lambda_s) = 
% \end{equation}
The ratio of the  $\Lambda$-parameters in any scale to $\Lambda_{\overline{\text{MS}}}$ can be determined by 
passing through the Wilson scheme as follows:  
\begin{equation}\label{eqn:convtoms}
    \frac{\Lambda_{\overline{\text{MS}}}}{\Lambda_s} \equiv \frac{\Lambda_{\overline{\text{MS}}}}{\Lambda_w}\frac{\Lambda_w}{\Lambda_s}\,,
\end{equation}        
where the separate factors can easily calculated in perturbation theory and are reported in Tab.~\ref{tab:improvedcouplings}.
Now using Eq.~\eqref{eqn:aofb} truncated to order $\lambda_s$ and fixing $\Lambda_s$ in terms of $\Lambda_{\overline{\text{MS}}}$ we can determine this quantity in units of $\sqrt{8t_1}$ from each of our values of $b$ and from all of the improved couplings. The result is displayed in  Fig.~\ref{fig:ascaling}. Obviously, a perfect asymptotic scaling would mean that all the values obtained were the same. However, it is natural to expect deviations coming from higher order perturbative and non-perturbative corrections to Eq.~\eqref{eqn:aofb}. These corrections should nonetheless vanish as we approach the continuum limit ($a=0$). This seems to be the case for our data since the larger difference obtained when comparing $\lambda_I$ with $\lambda_E$ approaches zero as we move towards smaller values of the lattice spacing. To give an estimate we perform a linear plus quadratic extrapolation in $a$ for each of the improved couplings separately displayed by the coloured bands in Fig.~\ref{fig:ascaling}.  The extrapolated values of $\Lambda_{\overline{\text{MS}}}$ in units of $\sqrt{8t_1}$ are $0.3582(47)$, $0.3620(46)$ and $0.3652(47)$ for $\lambda_I$, $\lambda_E$ and $\lambda_{E'}$, respectively. Notice that the three values are consistent within errors. The $\chi^2$ of the fits is not good but this is due entirely to the result at $b=0.38$ which is seen to deviate considerably from the others. We attribute this phenomenon to a bad estimate of the error based on the large autocorrelation times of the data at this value. A similar phenomenon has been observed and reported in similar studies of the flow in the literature~\cite{Hasenfratz:2023bok,Ishikawa:2017xam}. Our final results having good $\chi^2$ have been obtained by excluding the value at $b=0.38$, but the extrapolated values do not change significantly if we include it. 
A more detailed analysis of this behaviour will be covered in a future publication~\cite{paper}.

\begin{figure}
    \centering
    \includegraphics[scale=0.57]{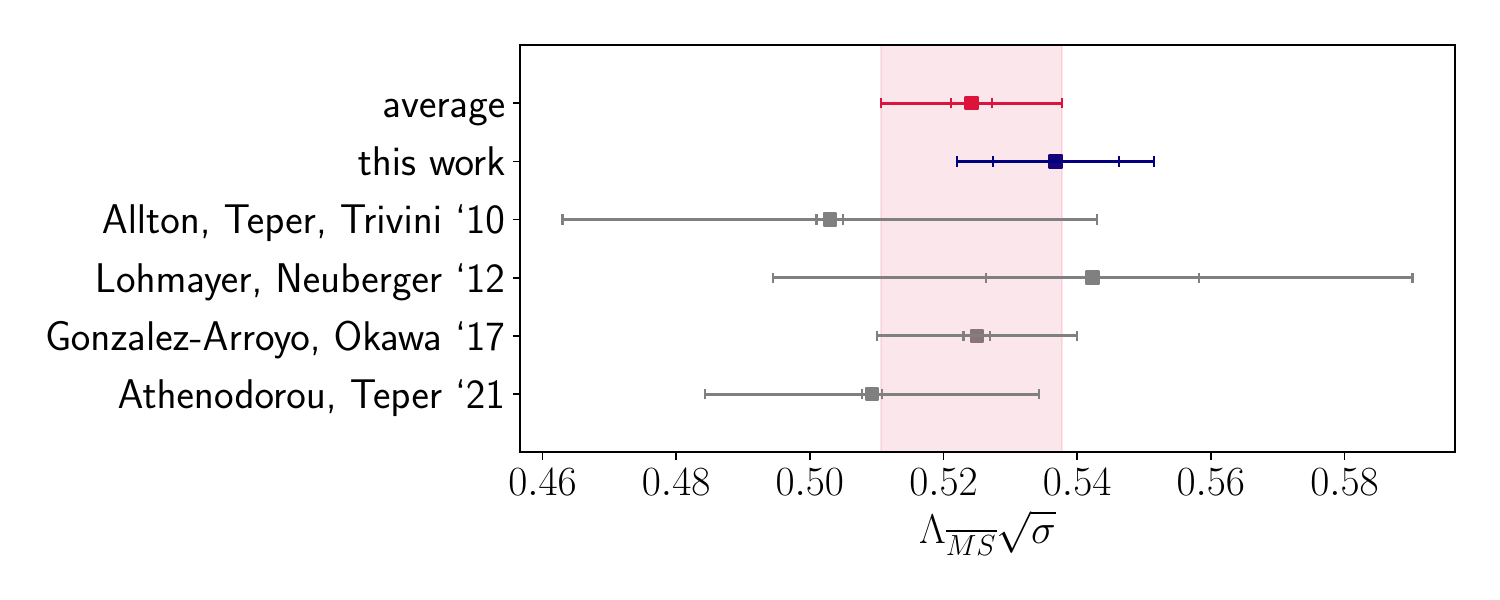}
    \caption{Values of $\Lambda_{\overline{\text{MS}}}$ present in the literature compared. The red band represents a weighted average.}
    \label{fig:lambda_flag}
\end{figure}

To quote a final estimate of the $\Lambda$-parameter for the Yang-Mills theory at large-$N_c$ in the $\overline{\text{MS}}$ scheme, we give the mean values between the previous results and assign the dispersion as a systematic error.
\begin{equation}
    \Lambda_{\overline{\text{MS}}} = 0.3618(47)(29) \, \qty(\sqrt{8t_1})^{-1} = 0.5366(94)(120)\, \sqrt{\sigma}\,,
\end{equation}
Notice that the use of the flow scale allows one to obtain an estimate at the level of 1-2 per cent. In the previous formula, 
we also report the final value in units of $\sqrt{\sigma}$ using the dimensionless ratio $R$ that we found in the previous section. For the systematic error, we have added in quadrature the systematic uncertainty on $\Lambda$ and the one on $R$.
Although this other estimate implies a loss in precision,  it has the advantage of allowing a comparison with previous determinations.  
These are $0.503(2)(40)\sqrt{\sigma}$ in~\cite{Allton:2008ty} and with the result $0.525(2)$ in the TEK model from~\cite{Gonzalez-Arroyo:2012euf} and also with the number of $0.5093(15)(250)$ given in~\cite{Athenodorou:2021qvs} and $0.542(16)(48)$ from~\cite{neu}. To collect and compare visually all these results we plot all the values in Fig.~\ref{fig:lambda_flag} in a FLAG-style plot.

%% file: body/3_conclusions.tex
\section{Conclusions}
We used volume reduction to study large $N_c$ Yang-Mills theory by simulating the TEK reduced model. This study focuses on testing predictions of asymptotic scaling for the evolution of the scale of the theory. We performed simulations for $SU(841)$ and for 8 values of the coupling and obtained a precise determination of the scale. Our study shows consistent results using different improved couplings, having small corrections and allowing a precise determination of the Lambda parameter of the theory in the $\overline{\text{MS}}$ scheme. Our work is part of a more extensive study to appear in a future publication~\cite{paper}.